\documentclass[12pt,preprint]{aastex}

\shorttitle{Sample Completeness and GRB Classification}
\shortauthors{Hakkila et al.}

\begin{document}


\title{How Sample Completeness Affects Gamma-Ray Burst Classification}


\author{Jon Hakkila and Timothy W. Giblin}
\affil{Department of Physics and Astronomy, College of Charleston,
    Charleston, SC, 29424}

\author{Richard J. Roiger and David J. Haglin}
\affil{Department of Computer and Information Sciences, Minnesota State
    University, Mankato, MN 56001}

\author{William S. Paciesas}
\affil{Department of Physics, University of Alabama in Huntsville, Huntsville, AL 35899}

\and

\author{Charles A. Meegan}
\affil{NASA, Marshall Space Flight Center, Huntsville, AL 35899}


\begin{abstract}
Unsupervised pattern recognition algorithms support the existence of three gamma-ray burst classes; Class I (long, large fluence bursts of intermediate spectral hardness), Class II (short, small fluence, hard bursts), and Class III (soft bursts of intermediate durations and fluences). The algorithms surprisingly assign larger membership to Class III than to either of the other two classes. A known systematic bias has been previously used to explain the existence of Class III in terms of Class I; this bias allows the fluences and durations of some bursts to be underestimated (Hakkila et al., ApJ 538, 165, 2000). We show that this bias primarily affects only the longest bursts and cannot explain the bulk of the Class III properties. We resolve the question of Class III existence by demonstrating how samples obtained using standard trigger mechanisms fail to preserve the duration characteristics of small peak flux bursts. {\em Sample incompleteness is thus primarily responsible for the existence of Class III.} In order to avoid this incompleteness, we show how a new dual timescale peak flux can be defined in terms of peak flux and fluence. The dual timescale peak flux preserves the duration distribution of faint bursts and correlates better with spectral hardness (and presumably redshift) than either peak flux or fluence. The techniques presented here are generic and have applicability to the studies of other transient events. The results also indicate that pattern recognition algorithms are sensitive to sample completeness; this can influence the study of large astronomical databases such as those found in a Virtual Observatory.
\end{abstract}


\keywords{gamma-rays: bursts---methods: data analysis, statistical---instrumentation: miscellaneous}


\section{Introduction}

In recent years, data mining algorithms have been used to aid the process of scientific classification. {\em Data mining} is the extraction of potentially useful information from data using machine learning, statistical, and visualization techniques. Pattern recognition algorithms (or classifiers) are data mining tools that search for clusters in complex, multi-dimensional spaces of {\em attributes} (observed and/or measured properties). These algorithms typically operate in one of two modes: supervised (in which the classifier is trained with known classification instances) and unsupervised (in which classification occurs without training examples).  Algorithms are designed to identify data patterns such as clustering and/or correlations, but their limitations must also be understood: it is up to the scientist to interpret physical mechanisms responsible for producing identified clusters.  Clusters found by classifiers can represent source populations; this happens when the class properties are produced by physical mechanisms pertaining to the sources. Clusters can also result from the way in which source properties are measured; sampling biases, systemic instrumentation errors, and correlated properties can all force data to cluster and thus give the appearance of class structure when there is none. 

Data mining algorithms can be applied to gamma-ray burst classification. Two gamma-ray burst classes have been recognized for years \citep{maz81,nor84,kle92,hur92,kou93} on the basis of duration and spectral hardness. Class I (Long) bursts are longer, spectrally softer, and have larger fluences than Class II (Short) bursts. Recent classification schemes have used data collected by BATSE (the Burst And Transient Source Experiment on NASA's Compton Gamma-Ray Observatory; CGRO) \citep{mee92} because this large database (2704 bursts observed between 1991 and 2000) was collected by a single instrument with known instrumental characteristics. Three attributes of BATSE gamma-ray burst data have been identified as being significant (using techniques such as principal component analysis) in delineating gamma-ray burst classes \citep{muk98,bag98,hak00}: duration T90 (the time interval during which 90\% of a burst's emission is detected), fluence S (time integrated flux in the 50 to 300 keV spectral range), and spectral hardness HR321 (the 50 to 300 keV fluence divided by the 25 to 50 keV fluence). Logarithmic measures of these values are typically used because classes are more clearly delineated when attributes are defined logarithmically. Historically, bursts with durations $T90 < 2$ seconds have been typically considered to belong to Class II. 

Data mining techniques allow a third gamma-ray burst class to be identified in BATSE data. Three classes are preferably recovered instead of two by both statistical clustering techniques \citep{muk98,hor98} and unsupervised pattern recognition algorithms \citep{roi00,bal01,raj02}. The third class forms at the boundary between Class I and Class II, and primarily contains the softest and smallest fluence bursts from Class I. Since Class II appears to be relatively unchanged by the detection of the third class, the three classes are called Class II (short, small fluence, hard bursts), Class I (long, large fluence bursts of intermediate hardness), and Class III (intermediate duration, intermediate fluence, soft bursts; also referred to as Intermediate bursts). 

The boundaries between classes are fuzzy, as some bursts are not easily assigned to a specific class. Different data mining algorithms do not necessarily assign individual bursts to the same classes because each classifier operates under different assumptions concerning correlations between data attributes and how these relate to clustering criteria. Some classifiers are designed to work with nominal data while others are not; some employ Bayesian while others employ frequentist statistics; some assume {\em a priori} distributions of attribute values while others do not. The results of any classifier can change if the size and makeup of the data set is altered. Data errors can influence the results since few classifiers currently include measurement error information in their analyses. However, irreproducibility is not necessarily a fault of machine learning methodology. Each classifier provides different insights into the way the data are structured.  For any given data set, there is a good possibility that some critical experiment or observation has not been performed, or that some key measurements have yet to be made, or that the relative importance of some attribute has been underestimated or overestimated. {\em There is no correct way of classifying a dataset because the usefulness of the classification depends on the insights gained from it by the user.}

In a previous application of supervised classification \citep{hak00} to gamma-ray burst data we hypothesized that Class III does not necessarily represent a separate source population. Instead, instrumental and sampling biases have been proposed as a way in which some Class I bursts can take on Class III characteristics. Due to a known correlation between hardness and intensity \citep{pac92,mit92,nem94,att94,dez97,qin01}, small fluence Class I bursts are typically softer than bright Class I bursts; this is supported by principal component analysis \citep{bag98}. Since the correlation results from a shift to smaller average peak spectral energy $\langle E_p \rangle$ at lower peak flux but not from changes in the average low-energy spectral index $\langle \alpha \rangle$ or the average high-energy spectral index $\langle \beta \rangle$ \citep{mal95,hak00,pac02}, this correlation has been attributed to the softer bursts being generally at larger cosmological redshift. (This conclusion may not necessarily be correct because a broad range of gamma-ray burst luminosities is suggested from redshifts of gamma-ray burst afterglows \cite{van00}; however, it should be noted that only a small afterglow sample is available.) Additionally, fluences and durations of some Class I bursts can systematically be underestimated \citep{kos96,bon97,hak00}; we refer to this as the {\em fluence duration bias} \citep{hak00}. Simply put, fluences and durations of some Class I bursts (particularly those with the smallest peak fluxes) can be underestimated due to the unrecognizability of low signal-to-noise emission; combined with their spectral softness, this gives them characteristics consistent with Class III.

Unfortunately, the fluence duration bias has been difficult to quantify. The amounts by which the fluence and duration of an individual burst are affected depend on the fitted background rates and estimated burst durations at all energies; to remove the background properly assumes {\em a priori} knowledge of the burst's intrinsic temporal and spectral structure. Such {\em a priori} knowledge can only be acquired in the absence of background, and gamma-ray burst observations are inherently noisy. Very high signal-to-noise estimates of a burst's temporal and spectral structure can only be obtained for a small number of the bursts with the largest fluences. These well-measured quantities are not entirely intrinsic; it appears that even the brightest bursts require systematic correction because they are at large redshift ($z \approx 1$).

Our objective is to determine whether or not the fluence duration bias can account for the number of bursts with Class III characteristics. In order to do this, we determine the total number of bursts that exhibit Class III characteristics using several different unsupervised classifiers. Then, we statistically model the suspected bias and determine whether it is strong enough to produce the Class III bursts. 

A number of pertinent questions will have to be addressed in pursuing this objective: Do theoreticians need to develop models for one, two, or three gamma-ray burst classes? How can data mining techniques be used to aid scientific classification? Are systematic effects present in data collected by BATSE or other gamma-ray burst experiments that alter classification structures? Can these effects be understood? Can information on intrinsic properties of the source population be extracted if these effects are present? Can future instruments be designed to minimize or eliminate these effects?

\section{Class III and the Fluence Duration Bias}

\subsection{The Significance and Size of Class III}

We systematically compare the output of various unsupervised algorithms in conjunction with a homogeneous gamma-ray burst data set obtained with one set of instrumental settings. We use the online gamma-ray burst ToolSHED \citep{hag00}  (SHell for Expeditions in Data mining) that we are developing to aid our analysis. This ToolSHED (currently ready for pre-beta testers at http://grb.mnsu.edu/grbts/) provides a suite of supervised and unsupervised data mining tools and a large database of preprocessed gamma-ray burst attributes. It allows users to classify data using more than one algorithm in order to identify consistencies in the different classification techniques and thereby gain better insight into the heterogeneous nature of the data. 

In order to further minimize the effects of instrumental biases, we have limited our database to bursts detected with a homogeneous set of BATSE trigger criteria. The database consists of bursts from the BATSE Current Burst Catalog (http://f64.nsstc.nasa.gov/batse/ grb/catalog/current/). Bursts included are non-overwriting and non-overwritten bursts ({\em e.g.} those whose BATSE readout periods did not overlap detectable bursts immediately preceding or following their detection) triggering at least two BATSE detectors in the 50 to 300 keV energy range with the trigger threshold set $5.5 \sigma$ above background on any of the three trigger timescales. We require all classifiers to use only the three attributes of $\log$(T90), $\log$(HR321), and $\log$(S). 

We apply four unsupervised ToolSHED algorithms with different approaches to clustering. These algorithms are ESX, a Kohonen neural network, the unsupervised EM algorithm, and the unsupervised Kmeans algorithm.

ESX \citep{roi99} is a classifier that forms a three-level tree structure. The root level of the tree contains summary information for all bursts. The second (concept) level of the tree sub-divides the root level into clusters formed as a result of applying a similarity-based evaluation function.  The third tree level holds the individual bursts.

A Kohonen neural network \citep{koh82} architecture is represented as a collection of input and output units.  During training, the input units iteratively feed the burst instances to the output units. The output units compete for the burst instances. The output units collecting the most bursts are saved. The saved  units represent the clusters found within the data.

The unsupervised EM algorithm \citep{dem77} assumes that the attribute space can be subdivided into a predetermined number of normally distributed clusters. An initial guess is made as to the properties of each random distribution, and this guess is used to calculate probabilities that bursts belong to each cluster. The cluster characteristics are iteratively adjusted until all clusters are optimally-defined.

The Kmeans \citep{llo82} algorithm randomly selects $K$ data points as initial cluster centers.  Each instance is then placed in the cluster to which it is most similar. Once all instances have been placed in their appropriate cluster, the cluster centers are updated by computing the mean of each new cluster.  The process continues until an iteration of the algorithm shows no change in the cluster centers.

Predetermined classification significance helps define the number of classes that can be recovered. When allowed to find an optimum number of classes based on a default significance, the aforementioned classifiers typically recover three to four burst classes as opposed to the two traditionally accepted classes.  This indicates that the two traditional classes are not considered to be the optimal solution.

We force all four classifiers to recover two, three, and four classes because we hope that by studying the properties of these force-recovered classes we can determine why the three-class solution is preferred over the two-class solution. The properties of three force-recovered classes are indicated in table~\ref{tbl-1}. The properties of these classes are similar to those obtained using other clustering techniques \citep{muk98,hor98,roi00,bal01,raj02}, so we again refer to these as Class I (Long), Class II (Short), and Class III (Intermediate). However, these previous results typically place fewer bursts in Class III than Class I, whereas three of our four classifiers place the largest number in Class III. Therefore, our analysis finds Class III to be the dominant class.

In order to explain why the percentage of Class III members is so large, we examine the placement of Class III bursts when classifiers are forced to recover only two classes (Short and Long). The results are remarkably consistent: all four classifiers fail to clearly delineate the traditionally-accepted Short and Long classes, and each places a large number of soft Class III bursts in with the hard Short class (see Figure 1). This is surprising, since Class III clustering is not obvious to the naked eye in the hardness {\em vs.} duration parameter space whereas Short and Long burst clustering is. The hardness {\em vs.} duration boundary is not chosen by the classifiers because a sharper one exists in the fluence {\em vs.} duration parameter space (Figure 2); {\em the boundary separating short faint bursts from long bright bursts is more significant than that separating short hard bursts from long soft bursts.}

It is surprising that fluence plays such an important role in the classification. First, fluence is an extrinsic attribute (since it represents a convolution of a burst's luminosity and distance) as opposed to hardness and duration, so there is no reason why fluence clustering should relate to any physical differences between burst classes. Second, one would intuitively expect a burst with a longer duration to have a larger fluence, indicating that fluence and duration should be highly-correlated attributes. Thus, clustering in the duration attribute can also cause clustering in the fluence attribute, and the use of fluence as a classification attribute magnifies the clustering importance of duration relative to hardness. The break between short faint and long bright bursts therefore appears due in part to the use of fluence, an attribute which is of questionable value. 

To determine if the fluence bias can be removed, we eliminate this attribute and perform the classification using only $\log$(T90) duration and $\log$(HR321) hardness ratio. Even without fluence, the classifiers again prefer to recover three classes instead of two, and Class III is not diminished in size. Examination of the three class properties indicates that $\log$(T90) has been used almost exclusively to delineate the classes; hardness is almost ignored by the classifiers. This is surprising, since the eye tends to delineate two burst classes. We check this result by supplying {\em only the T90 attribute} to the classifiers. Indeed, the classifiers again return three classes rather than two
(Class I bursts have T90 $> 6 $sec., Class II bursts have T90 $< 1.4$ sec., and Class III bursts have $1.4$ sec. $\leq$ T90 $\leq 6$ sec.). However, the size of Class III has been diminished in this reclassification and it is no longer the largest class; this result is consistent with that obtained earlier using only the duration attribute \citep{hor98}. We conclude that strong evidence exists for the three-class structure. 

Before accepting the new class as a separate source population, we must try to discount alternative explanations concerning its existence. It is possible that the classifiers have detected a data cluster resulting from the way that the data have been collected, rather than from a separate and distinct source population. We consider it unlikely that Class III represents a statistical anomaly since it has been found by four classifiers using different algorithms, and since stringent requirements have been imposed for each classifier to find additional classes. Thus, Class III could result from a systematic effect such as an instrumental or sampling bias. The suspected bias appears to primarily affect duration and the coupled yet extrinsic attribute of fluence.

This conclusion leads us again to examine the hypothesized fluence duration bias. This bias could provide a mechanism for underestimating both fluence and duration of some Class I bursts (particularly faint soft ones), and could cause these bursts to take on Class III characteristics. However, with the increased size of Class III, it is reasonable to think that the bias might be strong.

\subsection{Inadequacy of the Fluence Duration Bias Model to Explain Class III Properties}

In an attempt to quantify the fluence duration bias, we have developed a simple model of the bias that can be applied statistically. The model only influences Class I and Class III bursts (as defined by the EM algorithm), since the bias has not been hypothesized to alter Class II properties. In a previous work \citep{hak00} we estimated the maximum amount by which the fluences and durations of five bright bursts might need to be corrected if their signal-to-noise ratios were reduced; our simple model averages these values to obtain {\em maximum corrections} of fluence and duration as functions of $p_{1024}$ (peak flux measured on the 1024 ms timescale). We do not know how much the fluence of an individual burst might need to be corrected, therefore we assume that the fluence of each burst should be corrected between 0 ergs cm$^{-2}$ sec$^{-1}$ and the maximum fluence correction $S_{\rm max}$, and that the duration of each burst should be corrected between 0 seconds and the maximum duration correction T90$_{\rm max}$. The amount of the maximum correction is dependent upon the signal-to-noise ratio and thus on the peak flux; the suspected bias is more pronounced for bursts with peak fluxes near the detection threshold. We naively assume a probability $\rho$ that each burst's measured fluence and duration will be altered with equal probability in the intervals $[0,\log(S)_{\rm max}]$ and $[0,\log(T90)_{\rm max}]$. Thus, the modeled amount by which an individual burst's fluence would be affected by the bias is $\rho \log(S)_{\rm max}$ and the amount by which its duration would be affected is $\rho \log(T90)_{\rm max}$. The problem can be inverted to estimate how much observed burst fluences and durations have been underestimated as a function of $p_{1024}$.

If the fluence duration bias produces Class III properties, then (1) the faint Class I and Class III bursts (as measured by $p_{1024}$) should show evidence of having their fluences (and durations) systematically underestimated, and (2) no evidence of this bias should be present if this combined distribution has been properly corrected for the effect. We would thus like to compare both the observed distribution and the corrected distribution with the ``true'' distribution. Unfortunately, we do not know the ``true'' distribution.

If we assume that the bias has not affected the fluence and duration distributions of bright bursts (as measured by $p_{1024}$), then we can compare the corrected and uncorrected distributions of faint bursts to the observed distributions of bright bursts. The comparison can be made once we identify how the fluence and duration distributions scale with peak flux.

If a given burst's intensity were decreased (either by decreasing the burst's luminosity or if the burst were observed at a larger distance), then its fluence would decrease proportionally to its peak flux. This generic statement is false only in the presence of sampling and/or instrumental biases. The effect of time dilation due to cosmological expansion is an example of a sampling bias that can systematically affect fluence count rates differently than peak flux count rates. Since we measure the peak flux and fluence in the same energy channels, the primary source of bias is that the observed peak flux can be as little as $(1 + z)^{-1}$ of its actual value due to time dilation, whereas the fluence would not be expected to be lessened. This bias would cause the peak flux of distant bursts to be small relative to the fluence; note that this bias cannot explain Class III characteristics, since Class III bursts have fluences that are small relative to their peak fluxes. In going from bright bursts to faint bursts, a decreasing signal-to-noise ratio can cause fluences of faint bursts to decrease non-proportionally to peak fluxes; this is an example of a statistical (rather than systematic) instrumental bias.

Sampling biases can cause the faint burst distribution (as measured by either peak flux or fluence) to be different than the bright burst distribution. Trigger biases can cause bursts with certain characteristics to trigger disproportionately relative to other bursts. However, trigger biases that have been proposed prior to this manuscript do not appear to alter the makeup of the BATSE dataset by large amounts \citep{mee00}. {\em We therefore assume in testing the fluence duration bias that it is primarily responsible for causing a burst's fluence to change not in proportion to the change in its peak flux, and that the faint burst distribution of Class I + III bursts would be the same as the bright distribution in the absence of this effect.}

The distribution of burst fluences at a given peak flux indicates bursts with different time histories; greater fluence typically belongs to longer bursts with more pulses and smaller fluence typically belongs to shorter bursts with fewer pulses. If these burst peak fluxes were all decreased by the same amount, then their fluences would decrease proportionally along a line defined by $\log(S)_{\rm line} = \log(p_{1024}) + R$ (where $R$ is an arbitrary constant). The difference $\Delta \log(S) = \log(S)_{\rm line} - \log(S)_{\rm obs}$ obtained for each burst indicates the fluence offset of each burst from the line given its peak flux. The distribution of $\Delta \log(S)$ can be examined for bright bursts ({\em e.g.} those presumably unaffected by the bias) and for faint bursts ({\em e.g.} those affected by the bias). In the absence of any biases, the faint distribution will be similar to the bright distribution. If the fluence duration bias is present, then the faint distribution will differ from the bright distribution. The aforementioned statistical correction should make the ``corrected" fluence distribution $\Delta \log(S)_{\rm corr} = \rho \log(S)_{\rm max} - \log(S)_{\rm obs}$ more compatible with the bright distribution than is the uncorrected faint distribution. 

Figure 3 is a plot of $\log(S)$ {\em vs.} $\log(p_{1024})$ for the burst sample used in this study. Class I, II, and III bursts have been identified using the unsupervised EM algorithm. The proportional decrease of fluence and peak flux is shown for a hypothetical Class I burst (diagonal line); the curving path indicates how the bias might affect the measured fluence of this burst as a function of $p_{1024}$ (curving line) in the case where $\rho = 1$. The amount by which the fluence would need to be corrected $\Delta \log(S)_{\rm corr}$ is also shown (vertical line).

We construct eight $\Delta \log(S)$ bins for the set of bright bursts and eight bins for the faint bursts (the zero point for the $\Delta \log(S)$ scale is arbitrary, so we use $\Delta \log(S) = \log(S) - \log_{p1024} +6$). The dividing line between ``bright'' and ``faint'' bursts is set at $\log(p_{1024}) \geq 1$ photon cm$^{-2}$ sec$^{-1}$ since bursts brighter than this value should be essentially unaffected by the proposed fluence duration bias. The faint uncorrected $\Delta \log(S)$ distribution is moderately different than the bright distribution, with a $\chi^2 = 13.8$ for 7 degrees of freedom and a corresponding probability of $q = 0.055$. The fluence distribution (as determined from $\Delta \log(S)$) has been shifted to lower values consistent with the fluence duration bias.

In order to test the correction by the proposed model, we correct the fluence of each of the $i$ bursts by differing amounts $\rho_i \log(S_{\rm max})$. The $\chi^2$ of the corrected faint burst distribution is again compared to the ``control'' sample of bright bursts. Since we might have overcorrected some bursts while undercorrecting others, we run the analysis a total of 100 times and average the results. The corrected $\Delta \log(S)_{\rm corr}$ distribution is significantly different than the bright distribution (we obtain $\langle \chi^2 \rangle = 34.0$ for 7 degrees of freedom and a corresponding probability of $q = 2 \times 10^{-5}$ that the two distributions are identical) {\em indicating that our model has significantly overcorrected for the suspected bias.} Similar results are obtained using the $\Delta \log(T90)$ distributions.

Since we have apparently overestimated the amplitude of the fluence duration bias $\rho_i \log(S)_{\rm max}$ for typical bursts, we can decrease our estimate of the bias by introducing a free parameter D in the relationships $\rho_i D \log(S)_{\rm max}$ and $\rho_i D \log(T90)_{\rm max}$, where $D=0$ represents the uncorrected sample while $D=1$ indicates the originally hypothesized bias. In table~\ref{tbl-2}, we demonstrate the effectiveness of the fluence duration bias for different values of $D$ chosen arbitrarily. The model fit is only improved when we reduce our estimates of $\log(S)_{\rm max}$ and $\log(T90)_{\rm max}$ significantly.

We have shown previously \citep{hak00} that the maximum time interval used to calculate burst fluences decreases dramatically near the BATSE detection threshold (essentially no bursts in the 3B catalog with $p_{1024} <0.4$ photons cm$^{-2}$ sec$^{-1}$ have fluence durations $\ge 100$ seconds). This indicates that the fluence duration bias causes fluences and durations of very long bursts with small peak fluxes to be underestimated. Our current analysis supports this hypothesis: the $\Delta \log(S)$ and $\Delta \log(T90)$ distributions suggest that shorter bursts with small peak fluxes have probably not been affected by the bias, whereas some longer bursts have. Our experience with BATSE data analysis procedures is also in agreement: fluence duration intervals are rarely chosen to be shorter than many tens of seconds, and the time histories of only a few bursts are particularly susceptible to this bias \citep{kos96}. This should prevent a systematic bias from being introduced for short bursts with small peak flux but not necessarily for long bursts with small peak flux.

These results indicate that the fluence duration bias does not influence faint bursts to the extent hypothesized previously. The shorter Class I bursts, which were originally thought most likely to take on Class III characteristics via the bias, are apparently affected the least. The fluence duration bias appears to primarily influence the properties of some longer BATSE bursts. We conclude that the fluence duration bias is not responsible for the large number of shorter softer bursts comprising Class III.

\section{Sample Incompleteness and the Duration Distribution}

Although the fluence duration bias does not appear to be responsible for the creation of Class III, our analysis of the proportional decrease of fluence and peak flux has unexpectedly provided new insight into measured burst properties. The faint fluence and duration distributions used in classification are truncated because the samples triggered using short-timescale peak fluxes. This truncation has the potential of biasing the sample via sample incompleteness. In order to study the potential effects of sample truncation, we consider the advantages of a {\em fluence-limited} sample relative to a peak flux-limited sample. 

An experiment triggering with a short integration window is more likely to detect a short burst than an experiment triggering with a long integration window, because in the latter case the entire burst flux can be recorded in a single temporal bin. A peak flux-limited sample is thus biased towards shorter bursts relative to longer bursts because it excludes longer bursts having large fluences but with peak fluxes too faint to trigger. However, a long timescale trigger (such as one that could trigger on fluence) would prove to be equally-biased. A hypothetical experiment triggering on fluence ({\em e.g.} one with a 10,000 second integration window) would be more likely to detect a faint long burst (having little of its fluence in one temporal window) than would an experiment triggering on peak flux. A fluence-limited sample would be biased towards longer bursts because it would include faint longer bursts but exclude faint shorter bursts with the same peak flux. Figure 3 demonstrates that an excessive number of short Class III bursts are found near BATSE's peak flux trigger; the fluence distribution of these bursts is acutely truncated by the peak flux trigger. {\em Thus, Class III occupies a fluence {\rm vs.} peak flux region where the instrumental (peak flux) trigger favors detection of shorter bursts over longer ones.}

We would like to identify a peak flux measure that does not suffer from truncation of the duration distribution. The proportional relationship between fluence $S$ and peak flux $p_{1024}$ as a burst's luminosity is decreased or as its distance is increased provides a method for identifying such a peak flux measure. We can re-define fluence to be a peak flux by defining an extremely long temporal window $\tau$ ($\tau$ is a constant) that contains the entire flux of the sample's longest burst. The fluence divided by this temporal window $(S/\tau)$ is a peak flux (having units of photons cm$^{-2}$ sec$^{-1}$ or ergs cm$^{-2}$ sec$^{-1}$, using an approximate transformation of $A \approx 2.24 \times 10^{-7}$ ergs photon$^{-1}$) \citep{hak200}. The equation governing this proportional decrease in peak flux and fluence is
\begin{equation}
2 \log(\mathcal{F}_{\rm 0}) = \log(S/(A\tau)) + \log(p_{1024})
\end{equation}
or $\mathcal{F}_{\rm 0}^2 = S/(A\tau) (p_{1024})$ where $\mathcal{F}_{\rm 0}$ has units of flux and is thus a measure of burst intensity. We define this quantity as the {\em dual timescale peak flux} \citep{hak02} since it uses two different timescale measurements. The minimum value of the dual timescale peak flux $\mathcal{F}_{\rm 0}$ can be called the {\em dual timescale threshold}. The dual timescale peak flux is merely a multiple of this threshold value, $\log(\mathcal{F}) = \log(\mathcal{F}_{\rm 0}) + K$ (or $\mathcal{F} = K \mathcal{F}_{\rm 0}$), where $K$ is a constant. A dual timescale threshold can be defined as an instrumental setting for a gamma-ray burst experiment ({\em e. g.} by requiring $S/(\tau) (p_{1024})$ to exceed a trigger value), as a selection process on previously-detected events in a standard experiment, or with archival data from an experiment triggering independently on one temporal trigger at a time (such as BATSE). This latter concept is not new; several studies have developed their own {\em post facto} BATSE triggers using archival time series data \citep{sch99,kom01,ste01}.

The dual timescale peak flux treats longer bursts having larger fluences and smaller peak fluxes on an equal basis with shorter bursts having smaller fluences and larger peak fluxes: {\em these bursts with different temporal structures have something in common, which is that they have equal probability of detection using the dual timescale peak flux.} Their differences must therefore be defined by a line orthogonal to the dual timescale peak flux, {\em e. g.} one that satisfies the relationship
\begin{equation}
\log(\Gamma) = \log(S/A) - \log(p_{1024}).
\end{equation}
or $\Gamma = S/(A p_{1024})$ where $\Gamma$ has units of time and represents a duration. We call $\Gamma$ the {\em flux duration} $\Gamma$; it measures the total time that a burst could emit at its peak flux in order to produce its fluence. Longer bursts typically have large $S/p_{1024}$ values and shorter bursts should typically have small $S/p_{1024}$ values. In fact, the correlation for Class I + III bursts demonstrated in Figure 4 has a Spearman Rank-Order correlation significance 0f $10^{-118}$ that $\Gamma$ and T90 are uncorrelated. 

The dual timescale peak flux does not favor bursts of any duration (longer than the smallest 1024 ms integration window), whereas peak flux or fluence do by truncating the distribution and thereby favoring ``faint'' bursts ({\em e.g.} those near the threshold) of longer or shorter durations. Since the dual timescale peak flux does not truncate the duration distribution, we can say that the dual timescale peak flux-limited sample retains the duration characteristics of the sample by preserving the duration $S/p_{1024}$ relative to the intensity $(S/\tau) (p_{1024})$.
 
It was recognized soon after BATSE's launch \citep{pet94} that long-timescale triggers underestimated the intensities of shorter bursts and biased the sample. However, our analysis demonstrates (perhaps surprisingly) that short temporal timescale triggers would also bias the sample against longer bursts. 

Figure 5 demonstrates how BATSE's peak flux trigger influences the number of events placed in Class III relative to those placed in Class I. Shorter bursts (small $S/p_{1024}$; denoted by region `C') have been detected while faint longer bursts (large $S/p_{1024}$; denoted by `A') have been excluded by the trigger. A hypothetical fluence trigger allowing the faintest shorter bursts currently detected by BATSE to trigger would not resolve this problem: shorter bursts (large $S/p_{1024}$; denoted by `B') would go undetected by the fluence trigger relative to longer bursts (small $S/p_{1024}$; denoted by `D'). A dual timescale threshold is shown (diagonal dashed line) that favors neither longer nor shorter BATSE bursts. The threshold excludes most of the bursts previously identified as Class III because these have been favored by the one-second trigger window relative to longer bursts. It also excludes many Class II bursts which are both typically faint and shorter than the one-second trigger window.
 
We make a cut on our BATSE sample that is equally complete for both longer (large $S/p_{1024}$) and shorter (small $S/p_{1024}$) bursts and use this as our dual timescale threshold. This threshold follows the relation $\log(S) + \log(p_{1024}) = -6.5$, and has been chosen so that even the longest bursts with the largest $S/p_{1024}$ values are detected by BATSE's actual peak flux trigger. This is demonstrated by the diagonal dotted line in Figure 4, and corresponds to a dual timescale peak flux (via equation 1) of $\mathcal{F}_{\rm 0} = 0.048$ photons cm$^{-2}$ sec$^{-1}$ for $\tau = 617$ seconds (the T90 of the longest burst in the sample).

We wish to determine how the sample properties vary with dual timescale peak flux. We thus divide our sample of Class I + III into four subsamples containing similar numbers of bursts but with different dual timescale peak fluxes: bright bursts ($\log S + \log p_{1024} \geq -5.1$, or $\langle K \rangle = 7.49$), moderately bright bursts ($-5.8 \leq \log S + \log p_{1024} < -5.1$, or $\langle K \rangle = 3.34$), faint bursts ($-6.5 \leq \log S + \log p_{1024} < -5.8$, or $\langle K \rangle = 1.49$), and the faintest bursts ($\log S + \log p_{1024} < -6.5$, or $\langle K \rangle = 0.66$). The first three samples are ``brighter'' than the dual timescale threshold, the faintest sample consists of bursts fainter than the dual timescale threshold and is primarily composed of bursts from Class III. The bin sizes are chosen so that the three with $\mathcal{F} \geq \mathcal{F}_{\rm 0}$ contain similar numbers of bursts, and so that each bin contains enough bursts to constitute a reasonable statistical sample. 

We identify three flux duration intervals from the sample: longer bursts ($\log S \geq \log p_{1024} - 5.6$), middle bursts ($\log p_{1024} - 6.1 \leq \log S < \log p_{1024} - 5.6$), and shorter bursts ($\log S < \log p_{1024} - 6.1 $). The bin sizes are again chosen so that each bin contains similar numbers of bursts, and so that each bin contains enough bursts to constitute a reasonable statistical sample. Longer bursts as measured by the flux duration ($\langle \Gamma \rangle = 20$ seconds) are also long as measured by T90 ($\langle T90 \rangle = 71$ seconds); the same correlation holds true for middle bursts ($\langle \Gamma \rangle = 6.25$ seconds and $\langle T90 \rangle = 24$ seconds) and shorter bursts ($\langle \Gamma \rangle = 2$ seconds and $\langle T90 \rangle = 8$ seconds). The quantity ${\rm T90}/\Gamma$ is the burst emission time relative to the flux duration; this is the amount by which the actual burst emission time is stretched relative to the time interval during which the burst could have emitted at the peak flux rate. It is interesting to note that $\langle \log \Gamma \rangle \approx \log(\rm T90)^{0.6}$ for Class I + III bursts. Bursts with $\log \Gamma$ and $\log$(T90) values that do not closely follow this relationship have unconvential time histories (see Figure 4).

The attribute $\Gamma$ is closely related to GRB duty cycle \citep{hak200}. The duty cycle $\Psi$ measures the persistence of burst emission via the relationship
\begin{equation}
\Psi = \frac{S}{A \cdot \rm{T90} \cdot p_{64}}
\end{equation}
where $A$ is the average energy per photon and $p_{64}$ is the 64 ms peak flux.
A large duty cycle ($\Psi \approx 1$) indicates persistent emission whereas a small duty cycle $\Psi \approx 0$ indicates sporadic emission. Using equation (2), it can be seen that $\Psi \approx \Gamma/{\rm T90}$. Thus, a burst with a large $\Gamma/{\rm T90}$ ratio is persistent because it emits at a high rate for a long time relative to its total duration. 

We have previously shown that Class II bursts have larger values of $\Psi$ and harder spectral indices than Class I and Class III bursts\citep{hak200}, supporting the hypothesis that these short, hard bursts belong to a different source population. On the other hand, Class III bursts are generally softer than Class I but have similar $\Psi$ values; the properties of these two classes overlap considerably.

If our hypothesis is correct that Class I + III comprises one population, then we expect that $\mathcal{F}$ and $\Gamma$ will deconvolve complex relationships previously measured with the attributes $S$ and $p_{1024}$. Figure 6 demonstrates the relationship between HR321 and $\mathcal{F}$ for the combined Class I + III burst sample. A strong correlation exists between hardness and dual timescale peak flux for bursts of all durations; the hardness ratios are similar for all bursts of the same $\mathcal{F}$ regardless of whether T90 or $\Gamma$ is used. It is also seen that the faintest bursts in the sample (short bursts fainter than the dual timescale trigger) appear to extend this relation. This evidence supports our hypothesis that the bulk of the Class III bursts are short Class I bursts that have preferentially been detected by BATSE's short timescale trigger. 

If a corresponding sample of longer Class I bursts is detected (by having a lower peak flux trigger threshold and/or by having some bursts trigger on a longer timescale), then these bursts most likely would be as soft as the Class III bursts. We suggest that the FXTs (Fast X-ray Transients) found by BeppoSAX \citep{hei01} using an x-ray trigger and subsequently identified in BATSE data \citep{kip01} might be long soft bursts that previously escaped detection.

The slope of the $\log$(HR321) {\em vs.} $\log$(peak flux) relation is largest when $\log(\mathcal{F})$ is used as the peak flux measure as opposed to either $\log(S)$ or $\log(p_{1024})$; this is true regardless of whether the sample is peak flux-limited, fluence-limited, or duration-limited. This result is demonstrated in table~\ref{tbl-3}, where hardness {\em vs.} intensity correlations are examined using a Spearman Rank-Order Correlation test for the three different intensity measures: the 1024 ms peak flux $p_{1024}$, the fluence $S$, and the dual timescale peak flux $\mathcal{F}$. Small probabilities indicate strong correlations between spectral hardness and the peak flux measure. Spectral hardness (and presumably redshift) correlates better with the dual timescale peak flux than with any other peak flux measure, regardless of which measure is used to {\em select} the sample. Furthermore, the $\log$(HR321) {\em vs.} $\log$(peak flux) slope is essentially identical for burst samples of different T90 durations when $\log(\mathcal{F})$ is used; it does not appear that the same can be said when either $\log(p_{1024})$ or $\log(S)$ are used as a peak flux measure. {\em Thus, $\mathcal{F}$ appears to more easily deconvolve the attributes of hardness, duration, and peak flux than do either $S$ or $p_{1024}$}. We take this to indicate that $\mathcal{F}$ is a preferred peak flux indicator to $S$ and $p_{1024}$.

There are potentially far-reaching consequences to having $\mathcal{F}$ as a less-biased temporal flux measure. To date, essentially all statistical studies have used either $\log(p_{1024})$ or $S$ as intensity measures (e.g. $\log(N > S)$ {\em vs.} $\log(S)$, $\log(N > p_{1024})$ {\em vs.} $\log(p_{1024})$, $E_{\rm peak}$ {\em vs.} $\log(p_{1024})$). These studies are potentially biased because $S$ and $p_{1024}$ do not deconvolve the hardness intensity correlation as cleanly as does $\mathcal{F}$. Presumably, studies made using $S$ and $p_{1024}$ combine longer bursts measured at one value of HR321 with short bursts measured at another value of HR321. The use of $\mathcal{F}$ in future modeling endeavors might improve our understanding of gamma-ray burst properties better than do either fluence or peak flux.

We test our hypothesis that Class I bursts and Class III bursts belong to the same population by submitting all bursts in the original sample brighter than $\mathcal{F}_{\rm 0}$ to the EM algorithm for unsupervised classification, and using the attributes of $S$, T90, and HR321. The classifier preferably recovers six classes as opposed to three; the original three-class structure is lost as a result of the new trigger. Despite this, Class II is still easily identifiable even though it contains only 40 members (Class II bursts in the BATSE Catalogs thus appear to have been preferentially detected as a result of BATSE's short timescale trigger). The remaining bursts are placed in five classes with properties not recognizable as belonging to the original Class I or Class III. These classes may provide interesting additional insights into burst properties, but they warrant no further discussion here because they are not identifiable as the original burst classes.

Thus, strong reasons exist that the Class III cluster arises primarily from the shape of the attribute space defined by BATSE's peak flux trigger, and not from a separate source population. Our results support the hypothesis that Class III does not represent a separate source population. We have demonstrated that both fluence and duration are truncated by BATSE's peak flux trigger. The truncation effectively oversamples short bursts relative to long bursts. As a result of this truncation, the database contains an excess of faint, short (soft) bursts. The use of the dual timescale trigger supports the hypothesis that Classes I and III are really one continuous duration distribution with faint bursts being softer than bright bursts. The properties of this continuous distribution become somewhat ambiguous at low signal-to-noise, where the fluence duration bias alters burst properties. 

On the other hand, Class II appears to represent a separate source population from Class I \citep{hak00}. Neither sampling biases nor instrumental biases appear to be responsible for creating Class II characteristics from Class I bursts. However, it should be noted that BATSE's short trigger timescales have aided in the large detection rate of these short events. 

\section{Conclusions}

We have demonstrated that

\begin{enumerate}
\item Gamma-ray burst Class III does not have to represent a separate source population; it can be produced by the integration time of the instrumental trigger,
\item the fluence duration bias by itself, as modeled from a sample of high signal-to-noise bursts, is unlikely to be responsible for the existence of Class III.
\item Class III is likely produced by an excess of short, low fluence bursts detected by BATSE's short trigger temporal window.
\item The excess bursts can be eliminated via a selection process that is dual timescale peak flux-limited, rather than peak flux-limited or fluence-limited. 
\item The dual timescale peak flux measure resulting from this selection process appears to correlate better with hardness (and therefore with $E_{\rm peak}$ and redshift) than either peak flux or fluence. This adds support to the argument that dual timescale peak fluxes correct the temporal limitations introduced by using single timescale peak fluxes. Dual timescale peak fluxes can be established for many combinations of temporal measurements.
\end{enumerate}

The results found here are important to gamma-ray burst astrophysics as well as to the general problem of scientific classification. Data mining tools can help identify complex clusters in multi-dimensional attribute spaces. The tools are sensitive to clusters and data patterns, as evidenced here because they have allowed us to discover clusters produced artificially as a result of sample incompleteness. This sensitivity is advantageous, because a better understanding of instrumental response and sampling biases can be used to improve the design of future instruments. 

We note that sample incompleteness is generic and applies to the detection of any transient sources identified as the result of a temporal trigger. Examples of transient event statistics that might be biased by a temporal trigger include flare stars, soft gamma repeaters, x-ray bursts, and earthquakes.

However, the sensitivity of data mining tools can also cause problems. Data mining is central to the operation of planned Virtual Observatories, which will electronically combine data collected from a variety of instruments with a range of temporal, spectral, and intensity responses. Since sample incompleteness can cause a single instrument with one set of characteristics to find phantom classes, classes identified using multiple instruments should be interpreted cautiously. The instrumental responses of Virtual Observatory components will have to be accurately known in order for newly-identified classes to be recognized as separate source populations.

It is important to recognize that data mining techniques have their limitations. Principal component analysis has identified fluence, duration, and hardness as being critical gamma-ray burst classification attributes, while the trigger attribute of peak flux was not chosen. Data mining classifiers failed to recognize that attribute selection had removed the attribute that could have provided the most insight into the gamma-ray burst clustering structure.


\acknowledgments

We gratefully acknowledge NASA support under grant NRA-98-OSS-03 (the Applied Information Systems Research Program) and NSF support under grant AST-0098499 (Research in Undergraduate Institutions). We also thank James Neff and Robert Dukes for valuable discussions.




\clearpage

\begin{figure}
\plotone{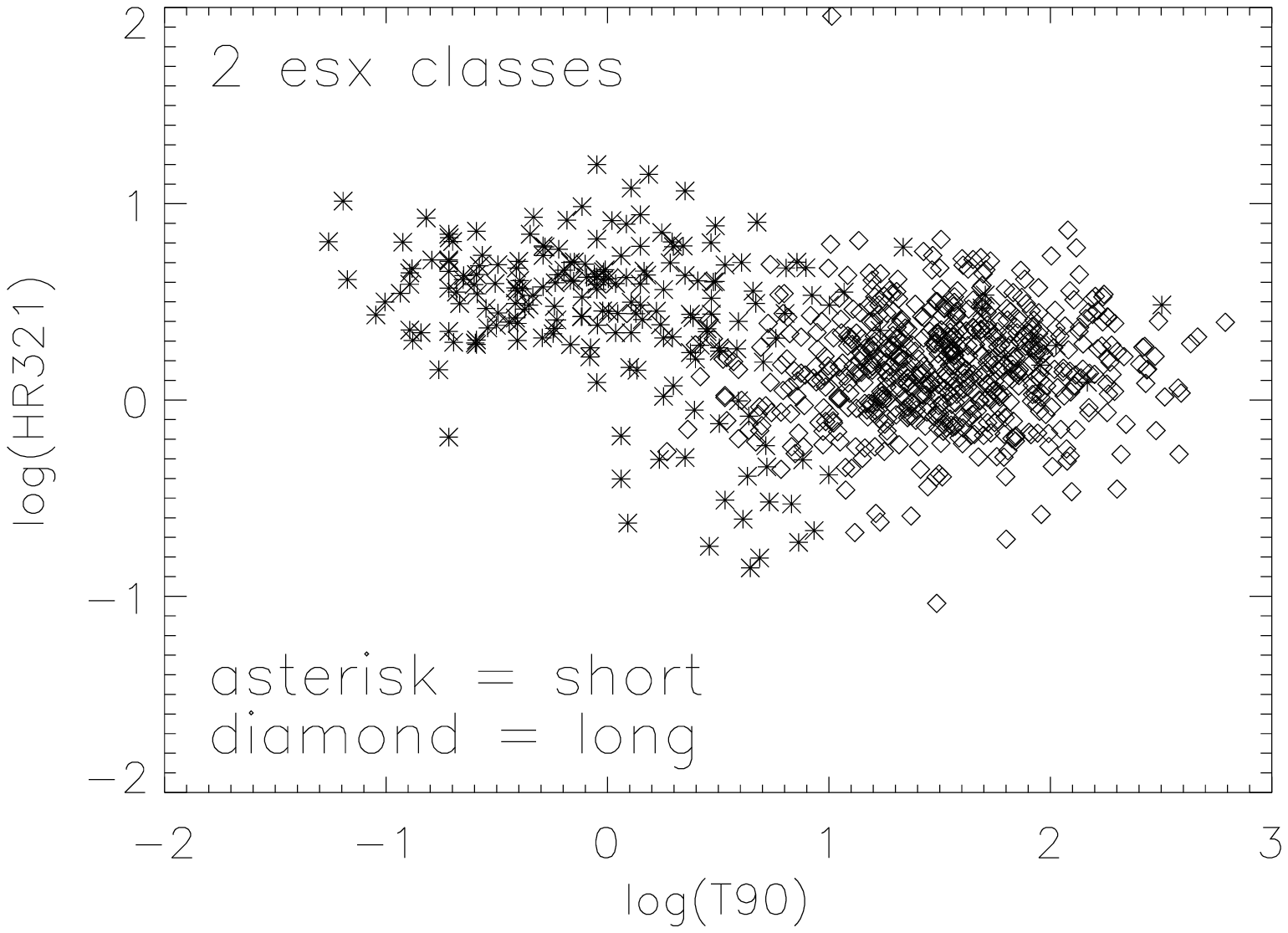}
\caption{Hardness and duration properties found by forcing the unsupervised classifier ESX to find two classes using 798 bursts in a sample defined by homogeneous trigger criteria. The two-class structure has forced many bursts traditionally placed in the Long class to be reclassified as Short. \label{fig1}}
\end{figure}

\clearpage 

\begin{figure}
\plotone{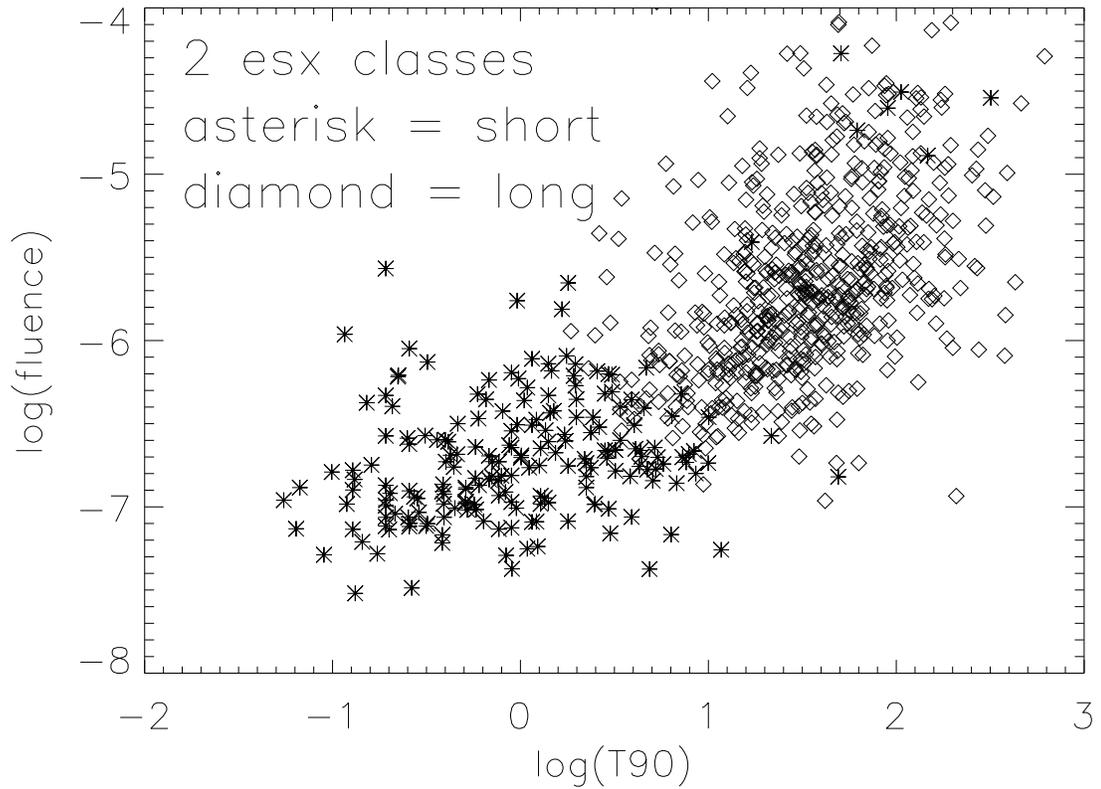}
\caption{Fluence and duration properties found by forcing the unsupervised classifier ESX to find two classes. A sharper division exists between classes in the fluence {\em vs.} duration parameter space than in the hardness {\em vs.} duration parameter space. Since fluence is an extrinsic attribute, we conclude that the fluence attribute is biased. \label{fig2}}
\end{figure}

\clearpage

\begin{figure}
\plotone{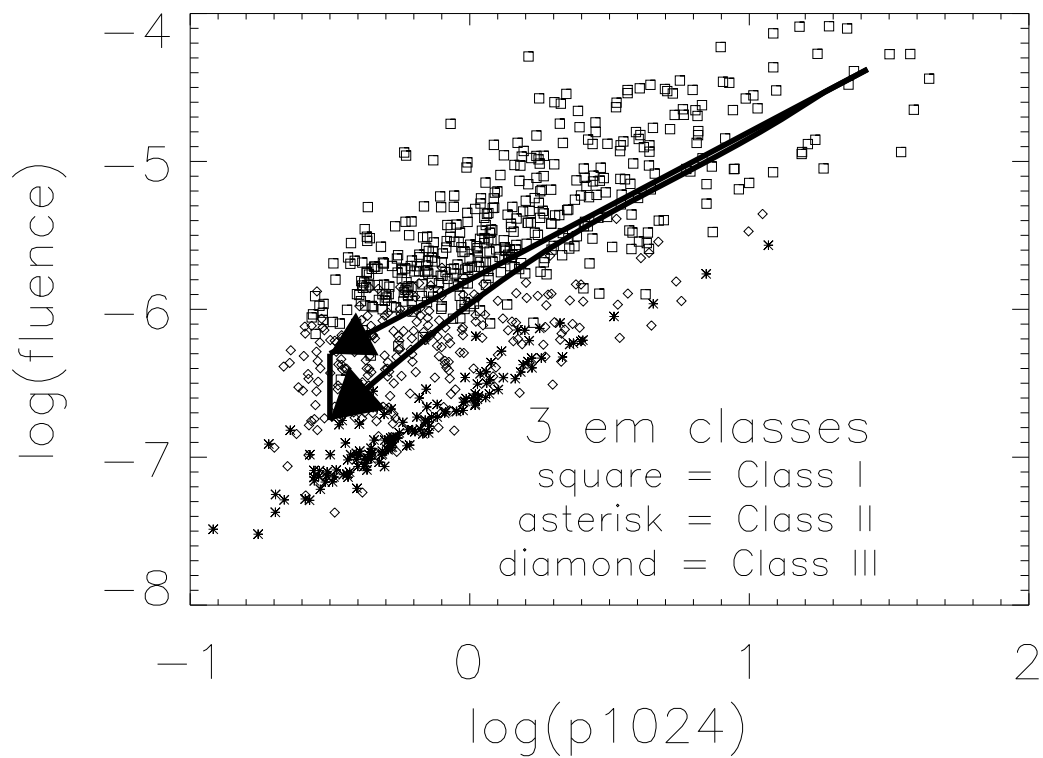}
\caption{Fluence vs. peak flux diagram showing the regions occupied by Class I (Long), Class II (Short), and Class III (Intermediate) bursts (as determined from the unsupervised EM algorithm). Maximum effects of the fluence duration bias are overlayed for a hypothetical Class I burst. The proportional fluence and peak flux decrease is shown (diagonal line) as is the maximum fluence decrease due to the bias (curving line). The maximum amount by which the fluence would need to be corrected $\Delta \log(S)_{\rm corr}$ is also shown (vertical line). \label{fig3}}
\end{figure}

\clearpage

\begin{figure}
\plotone{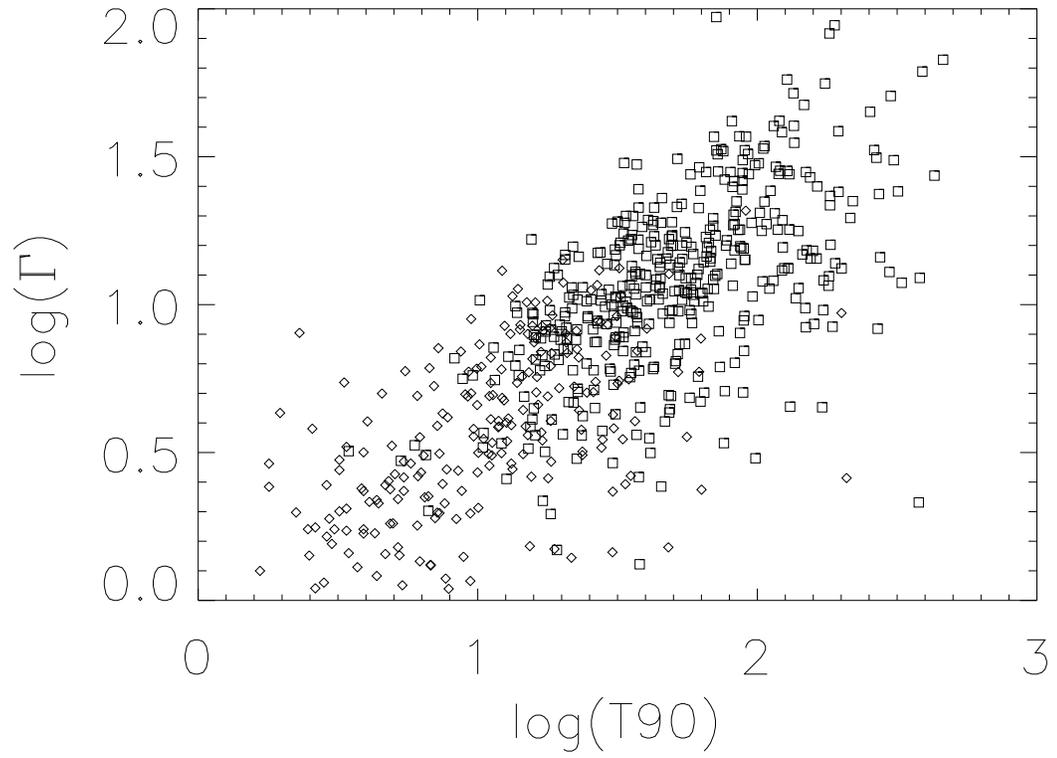}
\caption{Demonstration of the strong correlation between flux duration $\Gamma$ and duration $T90$ for the combined sample of Class III (diamonds) and Class I (squares). \label{fig4}}
\end{figure}

\clearpage

\begin{figure}
\plotone{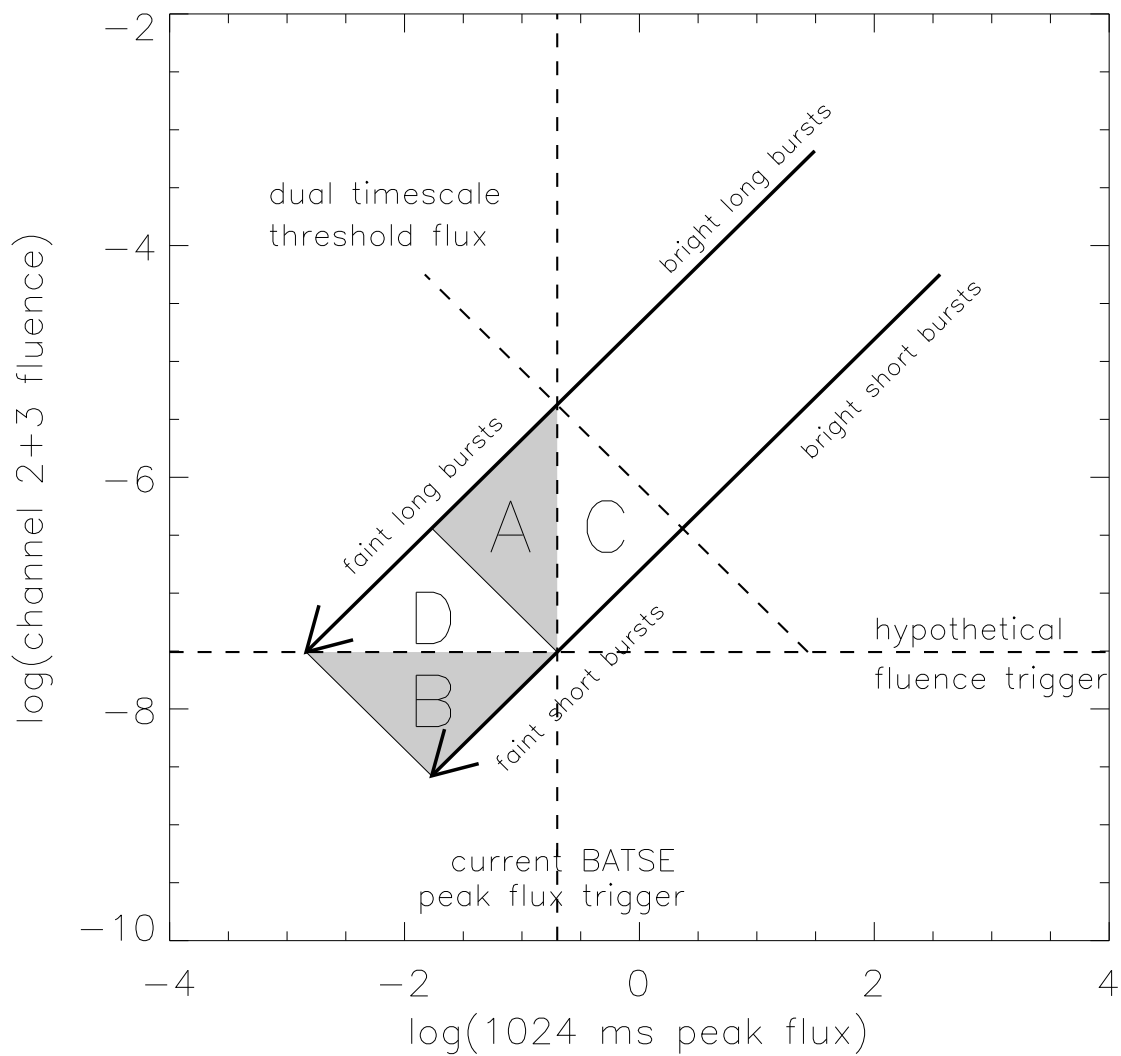}
\caption{Effects of fluence and peak flux triggers on selection of gamma-ray bursts. The BATSE 1024-ms trigger preferentially detects short bursts near threshold (region C), while missing longer bursts (region A). A hypothetical fluence trigger would preferentially detect long bursts (region D), while missing shorter bursts (region B). A proposed dual timescale threshold (which could be developed as an instrumental trigger on other experiments) would not oversample long or short bursts. \label{fig5}}
\end{figure}

\clearpage

\begin{figure}
\plotone{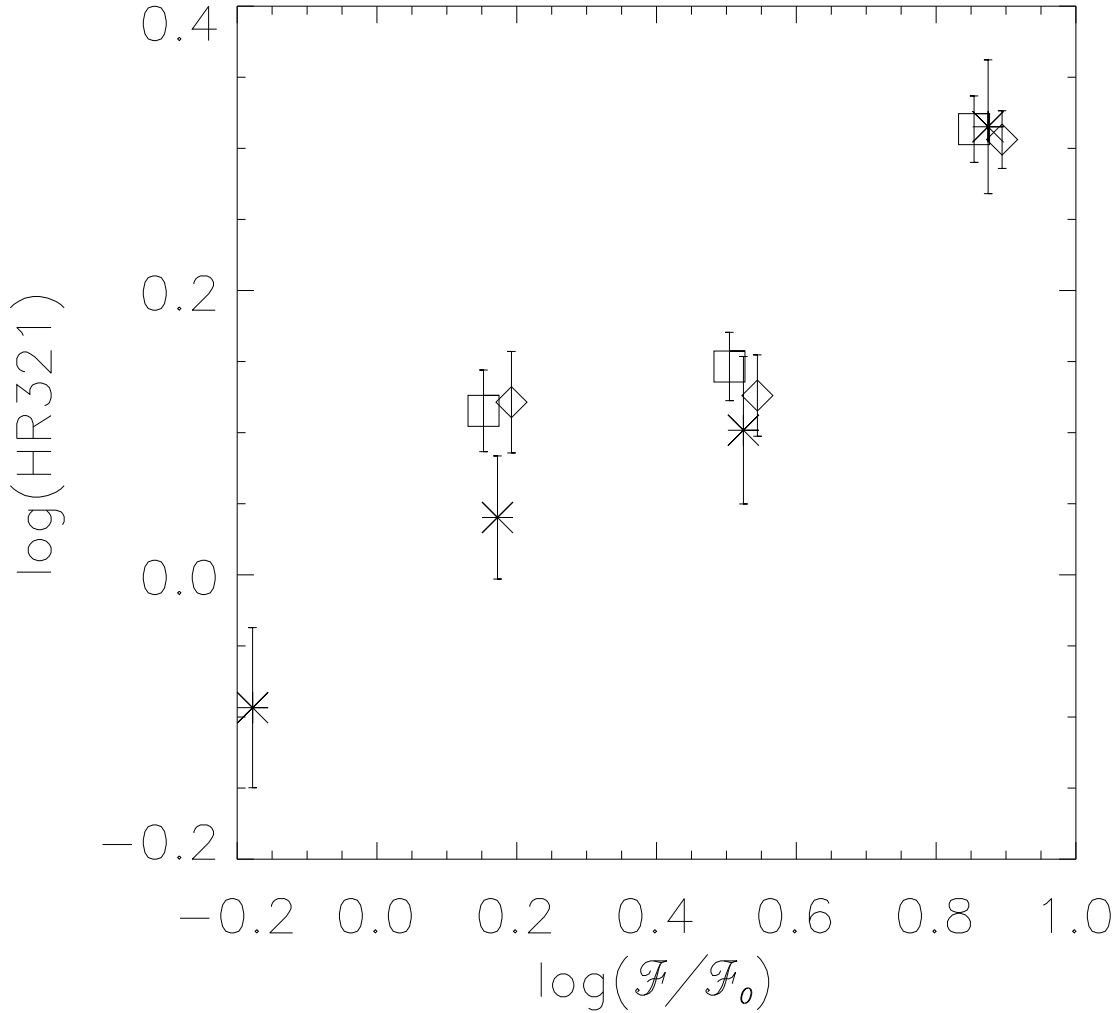}
\caption{Spectral hardness {\em vs.} dual timescale peak flux $\mathcal{F}/\mathcal{F}_{\rm 0}$ (normalized to the dual timescale threshold) for a binned sample of Class I + III bursts. 
The longest bursts (diamonds) have $\langle \Gamma \rangle = 20$ seconds, bursts of moderate duration (squares) have $\langle \Gamma \rangle = 6.25$ seconds, and shorter bursts (asterisks) have $\langle \Gamma \rangle = 2$ seconds.
Faint bursts (as measured by $\mathcal{F}$) are softer than bright bursts regardless of duration (long bursts are denoted by diamonds, bursts of moderate duration are denoted by squares, and short bursts are denoted by asterisks). The faintest bursts (short due to BATSE's short temporal trigger) are the softest of all. It appears that a faint sample of longer bursts (as measured by $\mathcal{F}$) should be as soft as the corresponding shorter bursts; these bursts require either a fainter peak flux trigger or a fluence trigger to be detected. \label{fig7}}
\end{figure}

\clearpage

\begin{table}
\begin{center}
\caption{Mean class properties when unsupervised classifiers are forced to recover three classes. Although each classifier produces different results, we refer to the similar recovered classes as Class I (Long), Class II (Short), and Class III (Intermediate). \label{tbl-1}}
\begin{tabular}{lrcccc}
\tableline\tableline
{\bf Class} & {\bf Property} & {\bf ESX} & {\bf Kohonen} & {\bf EM} & {\bf Kmeans}\\
\tableline\tableline
{\bf Class I} & {\bf No. bursts}  &   250   &    225     &  422  &  273\\
& {\bf $\log$(fluence)} & $-5.19$ & $-5.06$ & $-5.39$ & $-5.16$\\
& {\bf $\log$(T90)} & 1.71 & 1.71 & 1.70 & 1.78\\
& {\bf $\log$(HR321)} & 0.31 & 0.31 & 0.21 & 0.26\\
\tableline
{\bf Class II} & {\bf No. bursts} & 194 & 239 & 144 & 173\\
& {\bf $\log$(fluence)} & $-6.71$ & $-6.63$ & $-6.76$ & $-6.72$\\
& {\bf $\log$(T90)} & 0.02 & 0.13 & $-0.23$ & $-0.12$\\
& {\bf $\log$(HR321)} & 0.44 & 0.39 & 0.58 & 0.53\\
\tableline
{\bf Class III} & {\bf No. bursts} & 354 & 334 & 232 & 352\\
& {\bf $\log$(fluence)} & $-5.95$ & $-5.94$ & $-6.28$ & $-6.07$\\
& {\bf $\log$(T90)} & 1.38 & 1.51 & 1.02 & 1.28\\
& {\bf $\log$(HR321)} & 0.06 & 0.07 & 0.04 & 0.06\\
\tableline
\end{tabular}
\end{center}
\end{table}

\clearpage

\begin{table}
\begin{center}
\caption{Comparison of bright (large $p_{\rm 1024}$) burst distribution to the faint (small $p_{\rm 1024}$) burst distribution (which is presumed to be biased by the fluence duration bias). The fluence of each burst is ``corrected'' for the assumed bias by an amount $\rho_i D \log(S)_{\rm max}$ where $\rho_i$ represents a random probability that a burst has had its fluence underestimated by the bias and $D$ is an overall amplitude of the bias ($D=0$ indicates no bias and $D=1$ indicates a large bias). Each Monte Carlo model has been run 100 times and averaged, producing an average $\chi^2$, $\langle \chi^2 \rangle$ and a corresponding probability of exceeding $\chi^2$, $q$. Although the fluences of faint bursts appear to have been underestimated in a manner consistent with the proposed bias (based on the $D=0$ model), the amplitude of the bias is inconsistent with that originally proposed \citep{hak00}. The best fit amplitude ($q \approx 0.1$) is too small to account for the large number of faint bursts that have been placed in Class III. It also appears that bursts with large T90 values are more likely than those with small T90 to have had their fluences underestimated, supporting the hypothesis that the fluence duration bias does not entirely explain the existence of Class III. \label{tbl-2}}
\begin{tabular}{lccc}
\tableline
$D$ & $\langle \chi^2 \rangle$ & dof & $q$ ($> \chi^2$)\\ 
\tableline
1 & 34 & 7 & $2 \times 10^{-5}$ \\
0.667 & 23 & 7 & $2 \times 10^{-3}$ \\
0.5 & 19 & 7 & $10^{-2}$ \\
0.25 & 13 & 7 & $7 \times 10^{-2}$ \\
0.1 & 12 & 7 & 0.11 \\
0 & 14 & 7 & 0.055 \\
\tableline
\end{tabular}
\end{center}
\end{table}

\clearpage

\begin{table}
\begin{center}
\caption{Spearman Rank-Order Correlation probability that no correlation exists between hardness ratio HR321 and three different peak flux measures: the 1024 ms peak flux $p_{1024}$, the fluence $S$, and the dual timescale peak flux $\mathcal{F}$. Small probabilities indicate strong correlations between spectral hardness and the peak flux measure. The results indicate that spectral hardness (and $E_{\rm peak}$; therefore presumably redshift) correlates better with the dual timescale peak flux than with any other peak flux measure, regardless of which measure is used to select the sample. Larger probabilities are found for the $S$-limited and $\mathcal{F}$-limited samples than for the $p_{1024}$-limited sample because these have been produced by trunctating data originally collected using the BATSE $p_{1024}$-limited sample. Note that $S$ produces a smaller probability with HR321 than $p_{1024}$ for a $p_{1024}$-limited sample; this is because the softest bursts have the smallest $S$ due to the truncated shape of the sampled parameter space ({\em e. g.} region C in Figure 5). Similarly, $p_{1024}$ produces a smaller correlation probability than $S$ for a $S$-limited sample. \label{tbl-3}}
\begin{tabular}{lccc}
\tableline
Prob. of no correlation between HR321 and: & $p_{1024}$ & $S$ & $\mathcal{F}$\\ 
\tableline
$p_{1024}$-limited sample & $1.63 \times 10^{-14}$ & $1.07 \times 10^{-19}$ & $9.94 \times 10^{-20} $\\
$S$-limited sample & $2.58 \times 10^{-11}$ & $1.30 \times 10^{-10}$ & $1.54 \times 10^{-12} $\\
$\mathcal{F}$-limited sample & $3.14 \times 10^{-14}$ & $1.47 \times 10^{-17}$ & $3.33 \times 10^{-19}$ \\
\tableline
\end{tabular}
\end{center}
\end{table}

\end{document}